\documentclass{article}

\usepackage{arxiv}

\usepackage[utf8]{inputenc} 
\usepackage[T1]{fontenc}    
\usepackage{url}            
\usepackage{booktabs}       
\usepackage{amsfonts}       
\usepackage{nicefrac}       
\usepackage{microtype}      
\usepackage{lipsum}
\usepackage{graphicx}
\graphicspath{ {./images/} }

\usepackage{caption}
\usepackage{subcaption}
\usepackage{algcompatible}
\usepackage{multirow}
\usepackage{textcomp, gensymb}
\usepackage{url}
\usepackage{makecell}

\usepackage{cite}
\usepackage{amsmath}
\usepackage{amssymb}
\usepackage{graphicx}
\usepackage{textcomp}
\usepackage{xcolor}

\usepackage{listings}

\usepackage[utf8]{inputenc}

\usepackage{siunitx}
\usepackage{booktabs}
\usepackage{graphicx}
\usepackage{amsmath}
\usepackage{siunitx}
\usepackage{stfloats}
\usepackage{xcolor}
\usepackage{graphicx}
\usepackage{colortbl}
\usepackage{multirow}

\usepackage{makecell}

\usepackage{xstring}
\let\oldtexttt\texttt
\renewcommand{\texttt}[1]{\oldtexttt{\StrSubstitute[0]{#1}{.}{.\allowbreak}}}

\usepackage{tikz}
\usepackage{pgfplots}
\usetikzlibrary{fillbetween}
\usepgfplotslibrary{dateplot}
\pgfplotsset{compat=1.17}

\usepackage[eulergreek]{sansmath}
\pgfplotsset{
every x tick label/.append style={font=\scriptsize\sansmath\sffamily},
every y tick label/.append style={font=\scriptsize\sansmath\sffamily},
every axis label/.append style={font=\small\slshape},
every axis legend/.append style={font=\small},
}

\definecolor{bargreen}{rgb}{0.72, 0.88, 0.65}

\definecolor{sparkblue}{rgb}{0.46, 0.76, 0.71}

\makeatletter
\newcommand\resetstackedplots{
\makeatletter
\pgfplots@stacked@isfirstplottrue
\makeatother
\addplot [forget plot,draw=none] coordinates{(0,0) (1,0) (2,0) (3,0) (4,0) (5,0) (6,0) (7,0)};
}

\usepackage{listings}
\usepackage{courier}
\definecolor{codegreen}{rgb}{0,0.6,0}

\usepackage{hyperref}       

\title{Smart meter data processing: a showcase for simple and efficient textual processing}

\date{September 20\textsuperscript{th}, 2021} 					

\author{
  Miguel Ferreira, André Neves, Rodrigo Gorjão, Carlos Cruz\\
  Unicage Europe\\
  Lisbon, Portugal\\
  \texttt{\{miguel.ferreira,andre.neves,rodrigo.gorjao,carlos.cruz\}@unicage.com} \\
   \And
  Miguel L. Pardal \\
  INESC-ID, Instituto Superior T\'{e}cnico, Universidade de Lisboa\\
  Lisbon, Portugal\\
  \texttt{miguel.pardal@tecnico.ulisboa.pt} \\
}

\begin{document}

\maketitle

\begin{abstract}
    The increase in the production and collection of data from devices is an ongoing trend due to the roll-out of more cyber-physical applications. 
    Smart meters, because of their importance in power grids, are a class of such devices whose produced data requires meticulous processing.
	In this paper, we use Unicage, a data processing system based on classic Unix shell scripting, that delivers excellent performance in a simple package.
	We use this methodology to process smart meter data in XML format, subjected to the constraints posed by a real use case. We develop a solution that parses, validates and performs a simple aggregation of 27 million XML files in less than 10 minutes. We present a study of the solution as well as the benefits of its adoption.
\end{abstract}


\section{Introduction}
\label{sec:intro}

Files are universal. 
Although implementations may differ, all computational systems use files as part of their normal operation. 
Text files are the simplest data container, where one can save data in a flexible and convenient manner.
Data saved in text files can be easily moved around and directly used by most applications. The direct use of files, without a mediator other than the operating system, has the potential to lower the usage of system resources and the total processing time.
However, this has to be done in a way that still manages large amounts of data effectively. 

We propose the use of the Unicage tools\footnote{\url{https://unicage.eu/}} to develop a big data system based on text files.
These tools are a set of small, simple and highly efficient programs that can be combined in a modular way to build robust, yet flexible, data processing pipelines. 
These programs cover a wide variety of basic procedures such as string formatting, mathematical functions, and also common operations used in relational queries such as joins and selections. 
A detailed comparison of the approach proposed by Unicage against other alternatives can be found in \cite{moreira2018leanbench}, where a study is performed in the context of batch and query processing of Internet of Things data.

To showcase the simple and efficient textual data processing using Unicage, we present a \textit{smart meter} use case. Smart meters play a central role in smart grids, producing data that is fundamental for the management of the grid itself \cite{SmartGridReview}. 
This data needs to be efficiently processed, not only to produce outputs in a timely way, but also to check for possible errors or unexpected events in the grid.
To achieve these requirements, we followed the Unicage methodology to develop a system that processes the raw data from smart meters, in an efficient and simple way.

We structure this paper as follows: in Section~\ref{sec:background} we introduce Unicage and its methodology, along with some background regarding smart meters and smart grids. In Section~\ref{sec:use_case} we present the challenges of the smart meter use case. 
In Section~\ref{sec:approach} we present an approach to deal with such challenges using the Unicage tools. In Section~\ref{sec:evaluation} and~\ref{sec:discussion} we characterize and evaluate the performance of the proposed approach and assess its adequacy to a production system. Finally, in Section~\ref{sec:conclusion}, we present the conclusions.

\section{Background}
\label{sec:background}

For the majority of computer users, the fundamental data storage abstraction is the file.
Out of all the possible file types, the simplest one is the text file. 
These files may contain structured or unstructured data in a human readable form.  
Moreover, having data on text files gives one the possibility of using all the built-in operating system (OS) utilities to manipulate, process and format its contents. 
Unix-like systems~\cite{Ritchie73} ship with a plethora of such utilities that survived the test of time, being reliable and presenting good performance. 
It is in this context that Unicage is positioned.

\subsection{Unicage}
\label{subsec:unicage}
Unicage offers additional tools geared towards performance 
to cover some gaps in the toolbox of built-in OS utilities, 
still following the Unix philosophy~\cite{gancarz2003linux}. This paradigm ensures that systems can be customized and adapted to new needs according to the circumstances. 

\begin{figure} [!ht]
    \includegraphics[width=0.8\columnwidth]{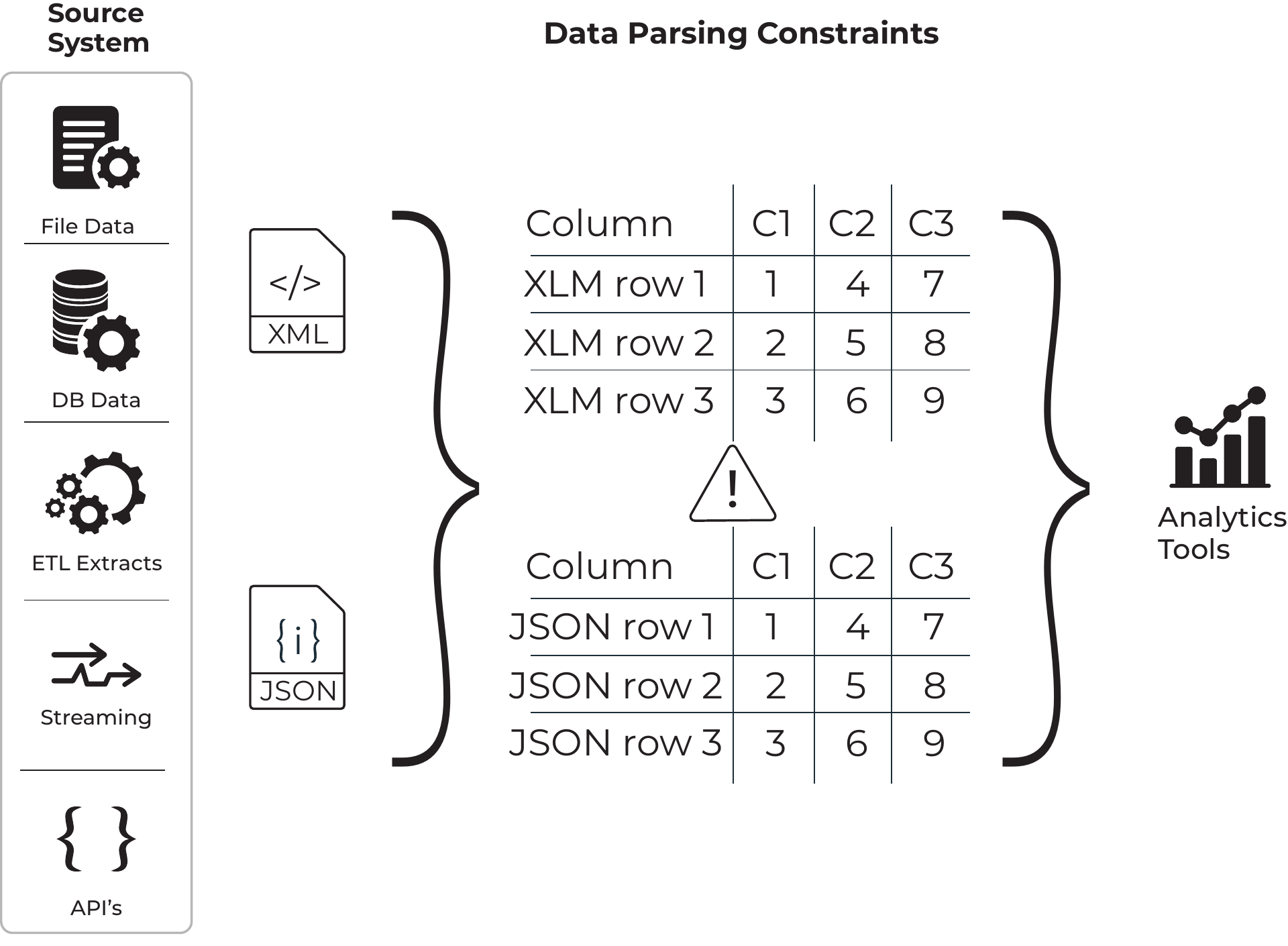}
    \centering
    \caption{Simple data parsing representation using Unicage. It is possible to acquire data from different sources and file formats, for example XML or JSON, and then parse it using the Unicage tools to obtain uniform and ready-to-use data for analytic tools, databases or other software.}
    \label{fig:Data Parsing}
\end{figure}

The ``glue'' that brings everything together is \textit{shell scripting}.
The data processing pipelines are composed of shell scripts that 
manipulate text files using a combination of Unicage tools and built-in OS utilities.
The different stages of data processing inside a shell script are connected through \textit{pipes}\footnote{A pipe is an  inter-process communication mechanism where the output of one command is connected to the input of the next command~\cite{robbins2004linux}.}, allowing the OS to split each stage in several processes. This, in turn, potentiates a better use of the system resources, namely, each process can be executed in a separate processor core, as available;
also, the process context switches are aligned with the flow of the data across the pipeline. 
Furthermore, the modularity of Unicage solutions allows them to be applied in 
multiple
situations associated with changing environments or with data in several formats, as illustrated in Figure~\ref{fig:Data Parsing}.

\subsection{Smart Meters}
\label{subsec:smart_meter_data_volume}
\begin{figure}
	\includegraphics[width=0.6\columnwidth]{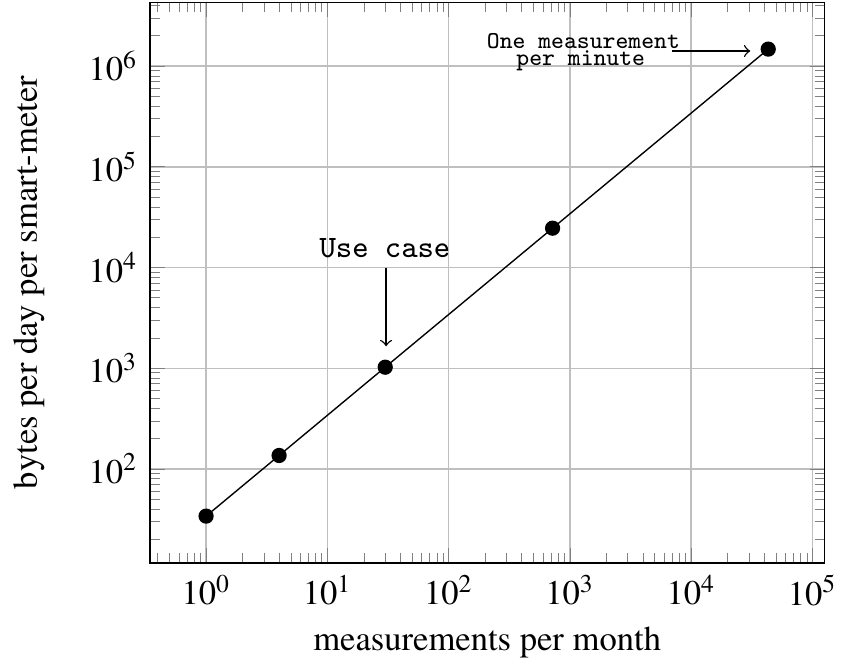}
	\centering
	\caption{Relation between the amount of data collected daily per smart meter and the frequency at which the readings are sent. As the number of readings collected 
	increases, so does the amount of data sent daily by each smart meter. In addition, we signal the amount of data considered in 1 reading per day (our use case), along with the case of 
	1 reading per minute 
	for each 
	smart meter.}
	\label{fig:projection}
\end{figure}

In recent years, there have been efforts by several companies and countries~\cite{edisonElectrical} to invest in the implementation of smart meters for energy infrastructures.
\textit{Smart grids}~\cite{SmartGridReview,Vincenzo2011SmartGP}
are a type of electrical power grid that allows two-way flows of electricity and information.
\textit{Smart meters}~\cite{s20102947} are devices that record and communicate information regarding the consumption of energy or other resources. 
Having smart meters in place allows the grid to automatically respond to conditions and events within it, which can lead to optimized energy distribution. However, for this to be possible, it is crucial to gather and disseminate data with strict time requirements~\cite{SmartMetersOverview}. 
This creates the need for a system that can process and manage large amounts of data efficiently.
The amount of data generated by smart meters depends on the frequency with which readings from the smart meters are collected. 
Supposing that:
\begin{itemize}
    \item each reading file has 1kB in size;
    \item each smart meter makes one reading per day;
    \item the grid has 27 million smart meters installed;
\end{itemize} 
one would have, approximately, 27GB of data produced every day by the smart meters. 
However, given that the frequency at which readings are collected tends to increase, so does the amount of data required to be processed, as can be seen in the projection shown in Figure~\ref{fig:projection}. 
Therefore, if we consider the same grid but increase the frequency to 1 reading per minute, then each smart meter produces 1,5MB per day; this means that, approximately, 38 TB of data are produced each day.

\section{Use case}
\label{sec:use_case}

Company T was collecting data from smart meters\footnote{This use case is based on a real challenge encountered by Unicage engineers. For confidentiality reasons, we do not use real names, data or infrastructure details; we present a stripped down version of the problem, keeping only its essential characteristics.}, once per month, from a total of 27 million devices (check Figure~\ref{fig:projection}). An upgrade in the grid infrastructure allowed the company to collect data once every day from each device, i.e., approximately 30 times more data to manipulate, organize and store. 
With this increase in the amount of received data, Company T decided it would be important to calculate daily metrics based on the readings
from the previous day. 
Company T also wanted to save the data from the readings in a database on the cloud. However, 
any reading that contained ambiguous data needs to be presented to a team of experts for review before being sent to the database. The validation step is important given that a wide range of decisions about the grid are made based on such data. The software solution the company had in place for the previous data demand was not capable of delivering the output in the required time limit for the new data volumes. Moreover, upgrading the hardware was not an option. The decision taken was to look for new software solutions that would process the data as efficiently as possible using the existing hardware.

In summary: Company T wanted a fast, reliable process that validated and processed the readings before sending them to the cloud for archival. For the use case, we also consider specific requirements: validate the readings based on their ``Reading Type'' and process the data to obtain the sum of the ``Value'' of the readings according to each ``Reading Type''.

\subsection{XML Files}
\label{subsec:xml-files}

The smart meters in the use case send the data in a custom XML format.
The eXtensible Markup Language~\cite{bray2000extensible} is a text-based format that allows the storage of data in an hierarchical structure. This structure is defined by tags that compose the elements of the structure. 
These tags are customizable and can be described in an XML-Schema with specified data types.

There are many\footnote{See, for instance, the ``Meter Data File Format Specification NEM12 \& NEM13'' from the Australian Energy Market Operator for more information} XML file standards for smart meter readings communication.
Listing~\ref{lst:xml} shows an example of the used format.
In our example, we have a set of tags that are used to identify the device\footnote{Most standards establish a much longer and more detailed section of this type, containing more information that characterizes the device.}, then we have a section that contains the readings themselves. Each reading, under the tag \texttt{Readings}, contains a timestamp, a value and a given type of measurement, which is identified by the 
\texttt{ReadingType}.

\begin{lstlisting}[label=lst:xml,caption=Smart meter readings XML file example.]
<MeterReadings>
    <MeterReading>
        <Meter>
            <Names>
                <name>SM000999VG</name>
                <NameType>
                    <description>This is a meter identification number.</description>
                    <name>MeterID</name>
                </NameType>
            </Names>
        </Meter>
        <Readings>
            <timeStamp>2021-03-08T22:22:18Z</timeStamp>
            <value>17.8280</value>
            <ReadingType ref="0.0.0.4.1.1.12.0.0.0.0.0.0.0.0.3.72.0"/>
        </Readings>
        <Readings>
            <timeStamp>2021-03-08T22:22:18Z</timeStamp>
            <value>17.9735</value>
            <ReadingType ref="0.0.0.12.1.1.37.0.0.0.0.0.0.0.0.3.38.0"/>
        </Readings>
        <Readings>
            <timeStamp>2021-03-08T22:22:18Z</timeStamp>
            <value>16.3959</value>
            <ReadingType ref="0.0.0.0.0.0.46.0.0.0.0.0.0.0.0.0.23.0"/>
        </Readings>
    </MeterReading>
</MeterReadings>
\end{lstlisting}

\section{Approach}
\label{sec:approach}

\begin{figure}
	\includegraphics[width=0.95\columnwidth]{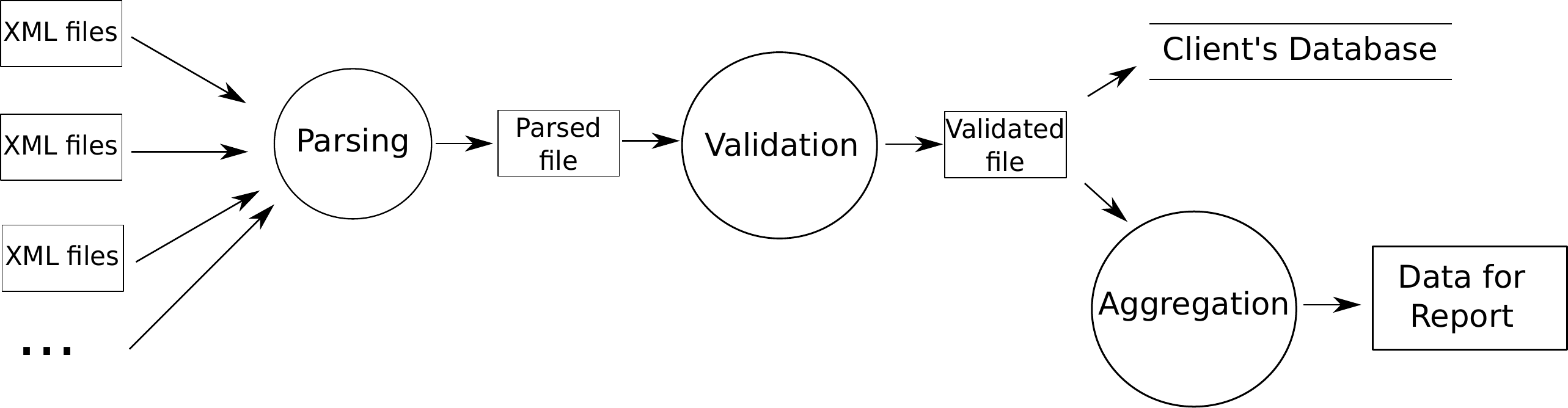}
	\caption{Data-flow representation of the solution comprising 3 scripts that perform the steps required by the use case: parsing, validation, and aggregation.}
	\label{fig:dataflow}
\end{figure}

Using the Unicage methodology, we developed 3 scripts that, combined, solve the problem. Each script performs a well defined task.
The whole process is represented in Figure~\ref{fig:dataflow}. These scripts were developed to be run after all the XML files have been received by the a dedicated machine. Once received, the XML files are fed to the parsing script (see Section~\ref{subsec:parsing}), where 
the resulting data is all merged into a single tabular file (see Listing~\ref{lst:parsed_file}). Then, the file 
is separated in valid and invalid data according to the reading code (see Section~\ref{subsec:validation}); the valid data is ready to be sent to client's database\footnote{The data upload procedure is out of the scope of the current solution.}. The valid data is also used to calculate the aggregated reading values (see Section~\ref{subsec:aggregation}) which will be made available to a subsequent process that will produce a report for the client. A detailed account of each step is made in the following sections, where we describe each script, summarizing their action and illustrating the most important pieces of code.

After running the scripts, the result is a set of plain text files with the data ready to be introduced in a database and a file containing the aggregated data to be used in a report.

\subsection{Parsing}
\label{subsec:parsing}

In a first step, we concatenate all the XML files and parse the resulting stream of data, separating it afterwards using the data characteristics. We achieve this by using the \texttt{find} command to identify all the XML files present in the given directory and then, using the \texttt{-exec} option, apply \texttt{cat} (concatenate) to all the identified files. This way, we can overcome the limited number of arguments imposed by \texttt{bash} when using the \texttt{cat} command. 

In a second step, we pass the output of the previous commands through a pipe -- the vertical bar in the end of the line -- to Unicage's \texttt{xmldir}. The \texttt{xmldir} command takes as input the hierarchy of the XML file that we want to parse and the file itself (in this case, the file is fed to \texttt{xmldir} through the pipe) and outputs a parsed version of the data in tabular form -- with an example shown in Listing~\ref{lst:xmldir_example}.

Then, we use a combination of Unicage's \texttt{self} (from ``\underline{sel}ect \underline{f}ield'') and \texttt{delr} (from ``\underline{del}ete \underline{r}ow'') with the \texttt{awk} command~\cite{aho1987awk}~
to prepare the data so that it can be transposed using Unicage's \texttt{map} command. Finally, we use Unicage's \texttt{delf} and \texttt{delr} to remove spurious data created as a side effect of the whole process. An example of the parsed data, ready to be validated, can be seen in Listing~\ref{lst:parsed_file}.


\begin{lstlisting}[label=lst:script1,caption=Script containing the parsing operations.]
find $readings_dir -name "*.xml" -exec cat {} +     |
xmldir /MeterReadings/MeterReading -                | 
self NF-1 NF                                        | 
awk '$1 == "name" || 
     $1 == "timeStamp" || 
     $1 == "value" || 
     $1 == "ref"'                                   | 
delr 2 "MeterID"                                    | 
awk '{if($1=="name")
	 {meter=$0;print $0}
     else
     {if($1 == "timeStamp")
     {print meter; print $0}
     else{print $0}}}'                              |
awk '{if($1 == "name"){count++}; 
     print count, $0}'                              |
map num=1                                           | 
delf 1                                              | 
delr 3 "0"                  > $parsed_dir/PARSED_FILE
\end{lstlisting}

\begin{lstlisting}[float=*t,label=lst:xmldir_example,caption=Example of the action of the xmldir command.]
[user@machine]$ xmldir /MeterReadings/MeterReading READINGS-SM000000001VG_20210101051308.xml 
MeterReadings MeterReading Meter Names name SM000000001VG
MeterReadings MeterReading Meter Names NameType description This is a meter identification number.
MeterReadings MeterReading Meter Names NameType name MeterID
MeterReadings MeterReading Readings timeStamp 2021-01-01T05:13:08Z
MeterReadings MeterReading Readings value 7.7190
MeterReadings MeterReading Readings ReadingType ref 0.0.0.4.1.1.12.0.0.0.0.0.0.0.0.3.72.0
MeterReadings MeterReading Readings timeStamp 2021-01-01T05:13:08Z
MeterReadings MeterReading Readings value 0.6193
MeterReadings MeterReading Readings ReadingType ref 0.0.0.12.1.1.37.0.0.0.0.0.0.0.0.3.38.0
MeterReadings MeterReading Readings timeStamp 2021-01-01T05:13:08Z
MeterReadings MeterReading Readings value 18.1170
MeterReadings MeterReading Readings ReadingType ref 0.0.0.0.0.0.46.0.0.0.0.0.0.0.0.0.23.0
\end{lstlisting}

\begin{lstlisting}[float=*t, label=lst:parsed_file,caption=First lines of the file produced by the parsing script.]
SM000000689VG 0.0.0.4.1.1.12.0.0.0.0.0.0.0.0.3.72.0 2021-01-01T12:40:06Z 14.8361
SM000000689VG 0.0.0.12.1.1.37.0.0.0.0.0.0.0.0.3.38.0 2021-01-01T12:40:06Z 7.4433
SM000000689VG 0.0.0.0.0.0.46.0.0.0.0.0.0.0.0.0.23.0 2021-01-01T12:40:06Z 6.5668
SM000000145VG 0.0.0.4.1.1.12.0.0.0.0.0.0.0.0.3.72.0 2021-01-01T08:54:15Z 19.7668
SM000000145VG 0.0.0.12.1.1.37.0.0.0.0.0.0.0.0.3.38.0 2021-01-01T08:54:15Z 10.1405
SM000000145VG 0.0.0.0.0.0.46.0.0.0.0.0.0.0.0.0.23.0 2021-01-01T08:54:15Z 6.9721
SM000000453VG 0.0.0.4.1.1.12.0.0.0.0.0.0.0.0.3.72.0 2021-01-01T06:50:54Z 9.9979
SM000000453VG 0.0.0.12.1.1.37.0.0.0.0.0.0.0.0.3.38.0 2021-01-01T06:50:54Z 19.0457
SM000000453VG 0.0.0.0.0.0.46.0.0.0.0.0.0.0.0.0.23.0 2021-01-01T06:50:54Z 14.0774
\end{lstlisting}

Notice that no transformation is performed on the data. The exact values that were received in the XML files are simply converted to a different format, which is easier to handle. 
The first column contains the ID of the meter, the second column the code identifying the reading type, the third column contains the timestamp and the fourth column contains the value of the reading. 

By combining all the data in a single file, we guarantee that the 
overhead associated with opening and closing a large number of files affects only the first stage of the process. All the subsequent scripts, namely the validation and the aggregation,
will deal with a single file.
Notice that these intermediate files  produced by the scripts can be deleted from the machine once the parsed data is sent to the cloud.

\subsection{Validation}
\label{subsec:validation}

The validation process is performed on the file generated by the parsing script. This file is filtered according to the data in the file \texttt{READING\_TYPE\_CONVERTER}, whose contents can be seen in Listing~\ref{lst:reading_types}. This file contains a mapping between the valid reading codes and the names of the readings. This index is given as input to \texttt{cjoin1} along with the data.

Generically, Unicage's \texttt{cjoin1} command aggregates data by index.
In this case, it looks through all the data in the parsed file and organizes it in two other files: the \texttt{VALID\_READINGS} file (see Listing~\ref{lst:validated_file}), which contains all the parsed data with a valid reading code (i.e. a code that was equal to one of the codes in the file \texttt{READING\_TYPE\_CONVERTER}), and the \texttt{INVALID\_READINGS} file, which contains the readings with an invalid reading code. The \texttt{INVALID\_READINGS} file is written through file descriptor 3, which is selected using the \texttt{+ng} option of the \texttt{cjoin1} command.  The \texttt{VALID\_READINGS} file contains an extra column with the name of the corresponding reading. This will be important for the aggregation script.

\begin{lstlisting}[label=lst:reading_types,caption=File containing a map between the reading types and their names.]
0.0.0.0.0.0.46.0.0.0.0.0.0.0.0.0.23.0 TYPE01
0.0.0.12.1.1.37.0.0.0.0.0.0.0.0.3.38.0 TYPE02
0.0.0.4.1.1.12.0.0.0.0.0.0.0.0.3.72.0 TYPE03
\end{lstlisting}

\begin{lstlisting}[float=*t, label=lst:script2,caption=Script containing the validation operations.]
cjoin1 +ng3 key=2 ./READING_TYPE_CONVERTER $parsed_dir/PARSED_FILE > $valid_dir/ALL_VALID_READINGS 
3> $tmp-ALL_INVALID_READINGS
\end{lstlisting}

\begin{lstlisting}[float=*t, label=lst:validated_file,caption=First lines of the produced by the validation script - ALL\_VALID\_READINGS.]
SM000000689VG 0.0.0.4.1.1.12.0.0.0.0.0.0.0.0.3.72.0 TYPE03 2021-01-01T12:40:06Z 14.8361
SM000000689VG 0.0.0.12.1.1.37.0.0.0.0.0.0.0.0.3.38.0 TYPE02 2021-01-01T12:40:06Z 7.4433
SM000000689VG 0.0.0.0.0.0.46.0.0.0.0.0.0.0.0.0.23.0 TYPE01 2021-01-01T12:40:06Z 6.5668
SM000000145VG 0.0.0.4.1.1.12.0.0.0.0.0.0.0.0.3.72.0 TYPE03 2021-01-01T08:54:15Z 19.7668
SM000000145VG 0.0.0.12.1.1.37.0.0.0.0.0.0.0.0.3.38.0 TYPE02 2021-01-01T08:54:15Z 10.1405
SM000000145VG 0.0.0.0.0.0.46.0.0.0.0.0.0.0.0.0.23.0 TYPE01 2021-01-01T08:54:15Z 6.9721
SM000000453VG 0.0.0.4.1.1.12.0.0.0.0.0.0.0.0.3.72.0 TYPE03 2021-01-01T06:50:54Z 9.9979
SM000000453VG 0.0.0.12.1.1.37.0.0.0.0.0.0.0.0.3.38.0 TYPE02 2021-01-01T06:50:54Z 19.0457
SM000000453VG 0.0.0.0.0.0.46.0.0.0.0.0.0.0.0.0.23.0 TYPE01 2021-01-01T06:50:54Z 14.0774
SM000000223VG 0.0.0.4.1.1.12.0.0.0.0.0.0.0.0.3.72.0 TYPE03 2021-01-01T03:08:28Z 14.4736
\end{lstlisting}

\subsection{Aggregation}
\label{subsec:aggregation}

The aggregation consists in a simple sum of the value of the readings according to their type. To do it, we start by isolating the relevant data from the \texttt{ALL\_VALID\_READINGS} file (see Listing~\ref{lst:validated_file}) using Unicage's \texttt{self} command. This way, we can select only the type of reading and the value, which correspond to columns 3 and 5, respectively. Then, using Unicage's \texttt{msort}, which implements the \textit{merge sort} algorithm, we sort the data by the reading type\footnote{Using  \texttt{msort} explicitly assures the use of \textit{merge sort}, independently of the underlying \texttt{sort} implementation available in the Linux distribution.}. This is required since the next command in the pipeline requires the data to be sorted. 

Finally, we use Unicage's \texttt{sm2} command to sum all the values by the reading type -- the first two arguments indicate the pivot of the sum (in this case the reading type is in column 1) and the last two arguments indicate the values to be summed. Listing~\ref{lst:aggregated_file} contains the result of this script. 

\begin{lstlisting}[label=lst:script3,caption=Script containing the aggregation operations.]
self 3 5 $valid_dir/ALL_VALID_READINGS          |
msort key=1                                     |
sm2 1 1 2 2 >  $corrected_dir/MEASURED_VALUES_by_TYPE
\end{lstlisting}

\begin{lstlisting}[label=lst:aggregated_file,caption=File resulting from the aggregation operation of  Listing~\ref{lst:script3}.]
TYPE01 10143.4150
TYPE02 9915.1172
TYPE03 10087.8375
\end{lstlisting}

\section{Evaluation}
\label{sec:evaluation}

The performance evaluation of the solution developed in Section~\ref{sec:approach} focuses on the execution time of the scripts. We measured the execution time of each script applied to a varying number of XML files, corresponding to smart meter readings. The results are presented in Figure~\ref{fig:results}. All the data points in the figure were calculated from a set of measurements;
we performed 40 steady-state time measurements and averaged those values. 
We consider \emph{steady-state} performance over \emph{start-up} performance, because the first is dominant in large enough files, once the processes are fully instantiated and loaded.
The maximum number of files used in the measurements is 1~million.
We did not perform tests with a larger number of files because, being anchored in the use case, we were able to build a system that was able to parse the 27~million XML files in batches of 1~million files.

\subsection{The hardware}
On the experiments, the tests were conducted on a computer with Intel Core i5-10210U CPU at 1.60GHz, 16 GB of RAM, 500GB SSD, running Ubuntu 18.04, Kernel version 5.4.0-77-generic and using build 2021/01/19 of Unicage.

\subsection{Baseline performance}

Since our approach is built on the process of transforming data saved in files, we use as baseline the performance of the copy command (\texttt{cp}) \footnote{\url{https://man7.org/linux/man-pages/man1/cp.1.html}}. To characterize the performance of the copy command, we made tests with a varying number of files. Because of the large number of files in the tests\footnote{\texttt{bash} imposes a limit to the number of arguments that can be passed to a command.}, we had to use the recursive option of the copy command (\texttt{cp -r}) so that, instead of copying file by file, we copied the whole directory containing the files.

\begin{figure}
    \centering
	\includegraphics[width=0.7\columnwidth]{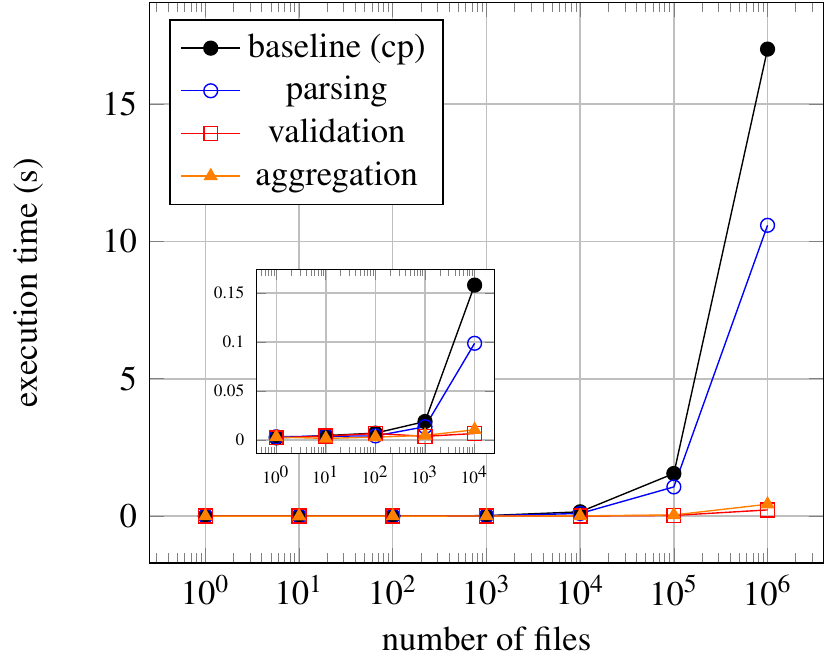}
	\caption{Comparison between the execution time of the different parts from our solution with the copy command (baseline)}
	\label{fig:results}
\end{figure}

\subsection{Processing stages performance}

The results of all the tests we performed can be seen in Figure~\ref{fig:results}. Two characteristics are clear from the plot:
\begin{itemize}
\item The performance of the parsing stage is the worst of all the processing stages, which is comprehensible since it deals with a large number of files directly;
\item The performance of the copy command is worse than the worst processing stage. This is related to the fact that, besides reading each file one by one, the copy command also has to allocate new places for them, one by one. The parsing stage reads each file individually too, but then it writes the results to a single file, allowing it to avoid the overhead of system calls associated with allocating a large number of individual files.
\end{itemize}

\section{Discussion}
\label{sec:discussion}

In the proposed solution the processing stages are executed in sequence, so the overall performance can be judged by combining the performance of each stage.
The three processing stages take around 12 seconds on average to complete for a batch of 1~million files: 11 seconds for the parsing, 0.3 seconds for the validation and 0.5 seconds for the aggregation.
In this case, 
we process
all the 27~million files in batches of 1~million. One could request the client to deliver the files in 
27~directories containing 1~million files each. According to these results, processing one directory at a time sequentially would take approximately 6 minutes
(27 times 12 seconds is 324 seconds, i.e. under 6 minutes).
This way, supposing that the data analysis process of the readings of a given day starts at 00h00 of the following day, this means that at 00h06 the database could start being loaded with validated values of the 27~million files and this data would be available to be included in the report that is presented at 07h00.

\subsection{Cloud Cost}
\label{subsec:cloud_cost}

The cost of running a complete operation on the cloud is affected by many factors. Next, we will focus only on the data storage.

Going back to the estimate that we mentioned in Section~\ref{subsec:smart_meter_data_volume}, if we apply it to the use case (i.e. 27 
~million smart meters) we see that the amount of data generated everyday is around 27~GB. This means that every month around 810~GB are produced.
%
We verified that the parsing operation to convert to a tabular format, just by itself, allows one to reduce 94\% of the storage space necessary to hold the data. This necessary step also has the beneficial side-effect of saving storage data, without loss of information. This means that the data saved on a set of XML files that needs 100~GB of space, can be stored in 6~GB in the parsed format. Applying this logic to the use case, this means that instead of sending 810~GB of XML files per month to the cloud, we could send 49~GB per month in parsed files.

\begin{figure}
    \centering
	\includegraphics[width=0.6\columnwidth]{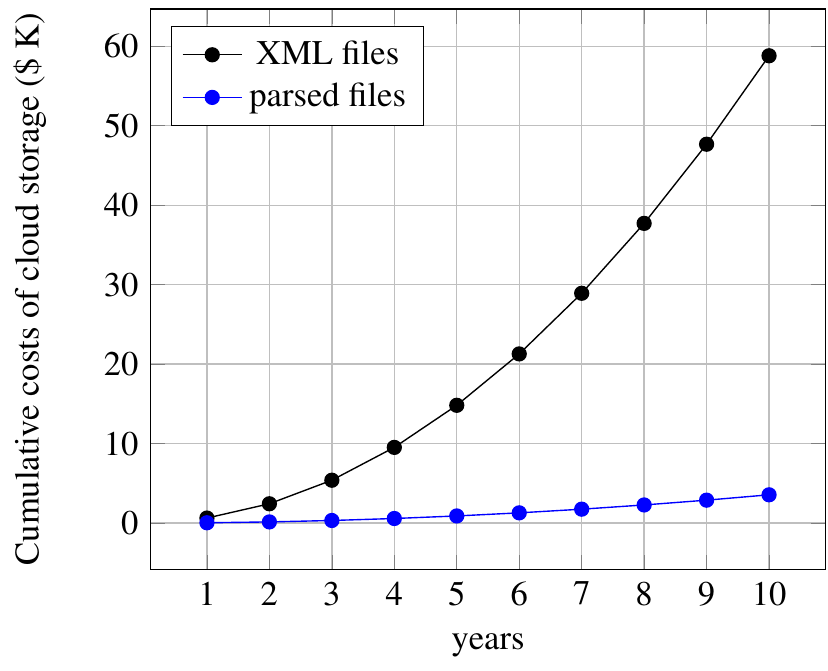}
	\caption{Estimate of cumulative costs of cloud storage per year. Using the notation of Eq.~\ref{eq:cumulative_cost}, $\alpha = \$ 0.01$, $D = 810$ for the XML files and $D = 49$ for the parsed data.
	}
	\label{fig:cloud_projections}
\end{figure}

Assuming there is a transfer of $D$ GB per month to the cloud, the total cost of the cloud storage service after $m$ months is given by
\begin{equation}
\label{eq:cumulative_cost}
C(m) = \frac{D \cdot \alpha \cdot m \cdot (m+1)}{2}, \quad m \geq 1
\end{equation}
where $\alpha$ is the cost per GB per month. For instance, at a rate of $D$ GB per month of new stored data, the storage service would cost, after a year, $78\cdot D\cdot \alpha$, where the currency is given by the value of $\alpha$; using the estimate of $D = 810$ for XML files, and the conservative approximation\footnote{Cheaper, tape-based storage solutions exist, but data is only available with a very large latency. In our context such solutions are not relevant.} of $\alpha = \$0.01$, the total cost after a year of service would be $\$ 631.8$. For the parsed data, $D = 49$, and the total cost after a year would be $\$ 38.22$. Figure~\ref{fig:cloud_projections} presents similar projection for longer periods of time using the conservative value of $\alpha = \$ 0.01$.

%
%
%
%
%

\begin{table}[h]
\begin{center}
\begin{tabular}{|l|c|}
\hline
 Number of XML Files & 27 million \\ \hline
 Average Processing time & 6 minutes \\ \hline
 Storage space reduction & 94\% \\ \hline
 Storage cost savings & 94\% \\ \hline
\end{tabular}
\end{center}
\caption{Main characteristics of the solution to the use case.}
\label{tab:summary_table}
\end{table}

\section{Conclusion}
\label{sec:conclusion}

The use of Unicage solutions for specific and customized tasks involving a large volume of data can be integrated seamlessly with any type of infrastructure. Being modular and versatile, such solutions can guarantee that data flows are more fluid between the nodes of a wider architecture, be it hybrid on-premises/cloud, cloud-only or fully on-premises.

We were able to construct a pipeline of shell scripts using a combination of the Unicage tools and built-in OS utilities to process large amounts of smart meter data from a real use case, as summarized in Table~\ref{tab:summary_table}. Our approach proved to be effective when dealing with big amounts of data, achieving a performance higher than the baseline defined for our tests. In addition, we show that by parsing the XML files and converting them to tabular text files, we were also able to considerably reduce their size, which led to a significant savings in cloud storage.




\bibliographystyle{unsrt}  
\bibliography{paper}

\end{document}